\documentclass[11pt]{article} 
\pdfoutput=1 

\usepackage{jheppub} 

\usepackage[utf8]{inputenc}
\usepackage[english]{babel}
\usepackage{amsmath,amssymb,amsbsy,amstext,amsthm,booktabs,simplewick,exscale,relsize,slashed,graphicx,amsfonts,upgreek, xcolor,subcaption}
\usepackage{multirow,color,dcolumn,bm,enumerate}
\usepackage{hyperref}
\usepackage{feynmp-auto,axodraw2,tikz}
\usepackage{ulem}
\usepackage{comment}

\usepackage{hyperref}
\usepackage[capitalise,noabbrev,nameinlink]{cleveref}

\title{Basis for Non-Factorizable Superamplitudes in $\mathcal{N}=1$ Supersymmetry}
\author[a]{Antonio Delgado,}
\author[a]{Adam Martin,}
\author[a]{Runqing Wang}

\affiliation[a]{Department of Physics, University of Notre Dame,
  South Bend, IN, 46556 USA}

\emailAdd{adelgad2@nd.edu}
\emailAdd{amarti41@nd.edu}
\emailAdd{rwang7@nd.edu}

\abstract{In this paper we develop a semi-standard Young tableau (SSYT) approach to construct a basis of non-factorizable superamplitudes in $\mathcal{N}=1$ massless supersymmetry. This amplitude basis can be directly translated to a basis for higher dimensional supersymmetric operators, yielding both the number of independent operators and their form. We deal with distinguishable (massless) chiral/vector superfields at first, then generalize the result to the indistinguishable case. Finally, we discuss the advantages and disadvantages of this method compared to the previously studied Hilbert series approach.}

\begin{document}
\maketitle

\setcounter{page}{2}

\section{Introduction}\label{intro}

In a previous work~\cite{Delgado:2023ogc}, we established a one-to-one correspondence between non-factorizable superamplidtudes, higher dimensional $N=1$ massless supersymmetry operators, and Young tableaux. The key to that connection is the existence of a $U(N)$ symmetry, where $N$ is the number of superfields in the operator/number of superstates involved in a superamplitude. Imposing this symmetry, the spinor helicity variables $\lambda_i, \tilde \lambda_i$ -- used to represent components of on-shell superstates -- are placed in the fundamental (for $\lambda_i$) and antifundamental (for $\tilde \lambda_i$) $U(N)$ representations. An additional ingredient when working with superamplitudes is the Grassmann coordinate $\eta_i$, which is used to construct superstates as (fermionic) coherent states. Invariance of the supercharges $Q_\alpha, \tilde Q_{\dot \alpha}$ demands that $\eta_i$ also transform as a fundamental of $U(N)$. Superamplitudes -- products of $\lambda_i, \tilde \lambda_i, \eta_i$ -- are thus tensor products of $U(N)$ fundamentals and antifundamentals, and have a natural home in terms of Young tableaux (YT). Crucially, kinematic constraints (on  $\lambda_i, \tilde \lambda_i$) such as equations of motion and integration by parts and supersymmetry constraints (on $\lambda_i, \eta_i$) from the Ward identities are manifest in the shape of the Young tableaux and make it easy to spot which combinations of $U(N)$ fundamentals and anti-fundamentals in the tensor product are viable. The net result is that, given a set of (massless, distinguishable) chiral/vector superfields and superderivaties, we get a unique Young tableau shape.

From that set of chiral/vector superfields and superderivatives, the next step is to form a operator basis. Specifically, how many independent operators are there with the specified field/derivative content, and what is their form -- meaning where are the derivatives applied and how are gauge and Lorentz indices contracted? If we are only concerned with the number of operators, Hilbert series techniques~\cite{Delgado:2022bho,Delgado:2023ivp} suffice (see Ref.~\cite{Lehman:2015via, Lehman:2015coa, Henning:2015alf, Henning:2017fpj} for a review of Hilbert series for non-supersymmetric effective field theories), however knowing both the number and the form is often more useful. For non-supersymmetric theories, higher dimensional operators can be represented by YT, and it has been shown~\cite{Henning:2019mcv,Henning:2019enq,Li:2020gnx} that labeling the Young tableau boxes with particle numbers, $1$ for the first particle in the operator, $2$ for the second, etc. and arranging the enumerated boxes according to the reduced semi-standard Young tableaux (SSYT) filling selects a basis. To make use of the SSYT technique, one needs to know how many boxes to fill for each field in the operator (= particle in the non-factorizable amplitude) -- how many $1$s, $2$s $\cdots N$s for an operator with $N$ fields. This number depends on the number of times $\lambda_1$, $\lambda_2$, etc. appear in the spinor helicity form of the operator. In Ref.~\cite{Henning:2019enq}, only a small number of operators were considered, and the authors just expanded out each operator of interest and counted the $\lambda_i$'s.  Subsequently, Ref.~\cite{Li:2020gnx} found a simple algorithm to determine the number of times a particular index $i \in 1\,..\, N$ appears purely from the number of derivatives present in the operator and the helicity of particle $i$ -- making it unnecessary to expand operators in spinor helicity form to determine the basis. With this simplifying step, the authors were able to determine the complete dimension eight operator basis in the Standard Model Effective Field Theory, and there have been several follow-ups to even higher mass dimension and expanded field content~\cite{Li:2020tsi, Li:2020xlh, AccettulliHuber:2021uoa, Li:2021tsq, Li:2022tec, Li:2023cwy,Song:2023jqm,Li:2023wdz,Harlander:2023psl,Harlander:2023ozs}. The first goal of this paper is to use the reduced SSYT filing to find a basis for superoperators, which boils down to find a similar algorithm for the number of times  index $i$ appears in YT representing superfield operators. This task requires altering the algorithm that works for non-supersymmetric theories, as now there is an additional ingredient, $\eta_i$, and there are two types of derivatives.

The second goal for this paper is to extend the operator $\leftrightarrow$ YT and SSYT techniques to operators involving indistinguishable fields. Indistinguishable fields introduce Bose/Fermi statistics, which need to be imposed by hand on the YT. This manipulation doesn't care about supersymmetry, and techniques for imposing Bose/Fermi statistics on YT have been discussed for non-supersymmetric theories in Ref.~\cite{Li:2020gnx}. We propose an alternative, and we believe simpler (especially for operators with fewer fields), technique for imposing identical particle symmetry/antisymmetry.

The rest of this paper is constructed as follows: In Sec~\ref{results}, we review the translation of superfield operators into YT form using a replacement rule that takes superfield operators to augmented spinor helicity variables. Next, in Sec.~\ref{sec:ssytbasis}, we introduce a  SSYT basis for YT. In order to apply the SSYT basis to YT from supersymmetric operators, we develop a counting scheme -- meaning how to translate the number and type of superfields involved in an operator to labels used to fill in the boxes of the YT. This scheme depends on the number of fields and derivatives alone. In Sec.~\ref{Indistinguishable Superfields}, we show how to systematically reduce the basis of operators when two or more fields are indistinguishable. The technique we use is independent of whether the operators are supersymmetric and is easily automated. We present our conclusions and a comparison of the YT and Hilbert series approaches to supersymmetric operator counting in Sec.~\ref{conclusions}.

\section{Methods and Results}\label{results}

The focus of our previous work, which we continue here, is to determine the number and basis for higher dimensional supersymmetric operators in a given class. By class, we mean a list of how many different chiral (and anti-chiral) and vector superfields are present, along with the number of derivatives, e.g. $D^2\overline D^2 \Phi^3 (\Phi^\dag)^2$. We will begin with the case when the fields are distinguishable (e.g. the $\Phi^3$ in the example just stated correspond to three distinct fields), returning later to indistinguishable fields. These higher dimensional operators can either reside in the superpotential (if purely chiral or anti-chiral) or in the K\"ahler potential. However as explained in Ref.~\cite{Delgado:2022bho}, terms with superderivatives can always be manipulated to sit in the K\"ahler term, so we will concentrate on setups of the form
\begin{align}
\int d^4\theta\, \mathcal{O}(\Phi, \Phi^\dag, W, \overline W; D, \overline D).
\end{align}
We assume that all fields are massless. Once the number of chiral and vector multiplets plus derivatives is fixed, the goal is to find the number of independent operators (free from equation of motion (EOM) and integration by parts (IBP) redundacies) and their explicit form.

In Ref.~\cite{Delgado:2023ogc}, we showed that each operator class can be represented by a Young tableau, at least when all fields are distinguishable. The shape of the Young diagram is set by the number of superderivatives and the number of chiral superfields present in the operator. In this section, we recap the derivation of the YT operator form. 

\subsection{From operator class to Young tableau}\label{YTform}

In order to  prove the YT form for supersymmetric operators, we followed the same logic as in non-supersymmetric theories. We  first introduced a replacement rule which takes Lagrangian level (off-shell) superfields to on-shell massless spinor helicity expressions. Spinor helicity expressions are on shell therefore they are functions of (super)states rather than of (super)fields. In supersymmetric theories, the replacements involve spinor helicity variables $\lambda_i, \tilde \lambda_i$ as well as $\eta_i$, a Grassmann variable introduced for each field in the operator to keep track of different components (helicity states) that are linked by supersymmetry,  for example  a superstate $\Phi_i = \psi_i + \eta_i\, \phi_i$ (a symbolic form for $|s = \frac 1 2 \rangle + \eta |s=0\rangle$, where $s$ is the helicity). Superstates can be formed in several different ways, depending on which supercharge one chooses to raise/lower helicity and whether one builds superstates starting from the highest helicity component (as in $\Phi$ above) or the lowest. Importantly, the choice of convention does affect how the supercharges act on the superstates and leads to some apparent asymmetry between how chiral vs. anti-chiral fields/states appear. In this work, we use the so-called $\eta$ convention/representation throughout. See Ref.~\cite{Delgado:2023ogc} for discussion of how to convert from one convention to another. 

The complete $\eta$ representation superfield replacement rules are listed below in Table~\ref{dicv} for chiral, anti-chiral, and vector superfields and their superderivatives.
\begin{table}[h!]
\begin{center}
\begin{tabular}{ |c|c|}
 \hline
Superfield & Spin-helicity Expression\\
 \hline
$\Phi_i $  & $\eta_i$   \\
$D\Phi_i $ & $\lambda_i$   \\
$\overline{D}D\Phi_i $ & $\tilde{\lambda}_i\lambda_i\eta_i$   \\
$\Phi^\dag_i $  & $1$   \\
$\overline{D}\Phi^\dag_i $  & $\tilde{\lambda}_i\eta_i$   \\
$D\overline{D}\Phi^\dag_i$ & $\lambda_i \tilde\lambda_i$\\
$W_i$ & $\lambda_i \eta_i$ \\
$DW_i$ & $\lambda_i \lambda_i$ \\
$\overline W_i$ & $\tilde \lambda_i$ \\
$\overline D\overline W_i $ & $\tilde \lambda_i \tilde \lambda_i \eta_i$ \\
$\overline{D}DW_i$ & $\lambda_i \lambda_i \tilde\lambda_i \eta_i$ \\
$D\overline{D}\overline{W}_i$ & $\tilde \lambda_i \tilde \lambda_i \lambda_i $\\ 
\hline
\end{tabular}
\end{center}
\caption{Replacement rule in the $\eta$ basis, expanded to include vector superfields. Additional powers of the superderivatives can be added to the above simply by applying $D = \lambda_i \partial/\partial \eta_i$ or $\overline D =  \tilde\lambda_i \eta_i$. Spinor indices have been suppressed, but can be reintroduced; note that when repeated $\lambda_i$ or $\tilde \lambda_i$ appear we are taking the symmetric spinor combination, e.g. $\lambda_i \lambda_i \to (\lambda_i \lambda_i)_{(\alpha \beta)}$. }
\label{dicv}
\end{table}
One may worry that the rules in Table~\ref{dicv} replace a field by its lowest component only, e.g. the chiral superfield $\Phi$ is mapped only to $\eta_i$, the coefficient of the scalar (lowest helicity) part of the superstate. However, this is completely compatible with the superspace formalism. To be more precise, the $\theta$ integration picks either the lowest component $\phi$ or the higher component $\psi$ and will never project out both two simultaneously. The replacement rule then states that the appearance of $\Phi$ in the expansion gives $\phi$, while $D\Phi$ gives $\psi$, etc. The latter identification is made between the lowest components of $D\Phi_{field}$ and $Q\Phi_{state}$, the fermion field $\psi$ in this case. In addition, massless EOM are automatically obeyed by replaced forms.

 Using the replacement rule, we derive the spinor helicity form for the operator class in question. This involves picking a particular partitioning of any derivatives, but any legal choice will suffice. We emphasize that using the replacement rule and picking an example operator from the class are crutches used for intermediate steps. Once we understand the YT form and how it depends on properties of the operator class, we can go directly from the operator class to the YT without replacing fields with spinor-helicity form.

Upon replacement, the representative operator is now a product of spinor helicity variables
\begin{equation}
\mathcal{O}(\Phi_i, \Phi^\dag_i, W_i, \overline W_i; D, \overline D) \sim f(\lambda_i,\tilde{\lambda}_i,\eta_i). \nonumber
\end{equation}
$\mathcal O$ is a Lorentz singlet, so all of the (suppressed in the above) spinor indices are contracted.\footnote{When gauge interactions are present, one expects the super gauge-covariant derivatives $\nabla_\alpha, \overline{\nabla}_{\dot\alpha}$ to appear rather than $D_\alpha$, $\overline D_{\dot\alpha}$. For the purposes of forming on-shell amplitudes/YT, what we care about is the Lorentz and supersymmerty properties of the derivatives, which are the same for $\nabla$ and $D$ (and $\overline{\nabla}, \overline{D}$). We will therefore use $D, \overline D$ for all derivative instances to keep things simpler.}

Rather than keeping track of these contractions, we endow the spinor helicity variables with $U(N)$ representations, where $N$ is the number of fields present in the operator; $\lambda$ and $\eta$ are taken to be $U(N)$ fundamentals, while $\tilde \lambda$ are taken to be antifundamentals\footnote{This assignment keeps the supercharges $Q = \sum_i \lambda_i \partial/(\partial \eta_i) , \tilde Q =  \sum_i \tilde\lambda_i \eta_i$ $U(N)$ invariant, and can be thought of as the extension of the little group for $N$ massless fields from $U(1)^N \to U(N)$. To keep track of fundamental vs. antidundamentals, we will use lower $i$ indices for the former and upper indices for the latter.}. This is useful, as the operator is now a tensor product of $U(N)$ fundamentals and antifundamentals and can be visualized using YT. Furthermore, the symmetries in a YT (whether boxes are symmetric or antisymmetric under interchange) are tied to the Lorentz (spinor) properties of the $\lambda, \tilde \lambda$. For example, in the spinor product $\epsilon_{\alpha\beta}\lambda^\alpha_i \lambda^\beta_j$ the spinor indices are antisymmetric. The $U(N)$ indices go along for the ride and are also antisymmetric, meaning the YT boxes for $\lambda_i$,$\lambda_j$ sit on top of each other as their own column of height two. Applying this logic to the set of $\lambda_i, \tilde \lambda_i$ in $\mathcal O$, the Lorentz singlets will be the tensor products of paired $\lambda_i, \lambda_j$ (columns of height two), and all pairs of $\tilde \lambda^i \tilde \lambda^j$ (antisymmetric combinations of antifundamentals). To avoid having to keep track of both fundamental and antifundamental indices, we convert each pair of antifundamentals into an antisymmetric product of $N-2$ fundamentals using the $U(N)$ epsilon symbol, $\tilde\lambda^i \tilde\lambda^j \to \epsilon_{1,\cdots i,j \cdots N}\tilde\lambda^i \tilde\lambda^j$. Diagram-wise, these appear as columns of height $N-2$ in the YT.

 We are now left with a collection of height two columns and a collection of height  $N-2$ columns. The next step is to  determine which tensor products between these two collections are legal. Additionally, we need to include the $\eta_i$, which we have said transform as $U(N)$ fundamentals, and we need to account for the integration over $d^4\theta$ to convert the K\"ahler term into a (higher dimensional) Lagrangian term. 
 
 As shown in Ref.~\cite{Henning:2019mcv,Henning:2019enq}, keeping only the YT with height $N-2$ columns to the left of the height two columns is equivalent to removing all IBP redundancies. We'll refer to this as the ``harmonic" YT form. Ref~\cite{Henning:2019mcv,Henning:2019enq} dealt with spinor helicity representations of non-supersymmertic theories. The fact that this result holds in our case -- for the $\lambda_i, \tilde \lambda^i$ pieces of the operator {\it before} integrating over $d^4\theta$ and ignoring the $\eta_i$ -- is because the Poincar\`e algebra is a sub-algebra of supersymmetry and the usual IBP redundancies only care about the total derivative generated by the same momentum operator $P$. In other words, a ``complete" YT diagram (which respect supersymmetry algebra and is IBP($\partial$)-free) contains a ``harmonic" YT as its sub-diagram. 
 
 The role of $d^4\theta$ and the $\eta_i$ dependence can be determined from the supersymmetric Ward identities. First, one can express $d^4\theta = \overline D^2 D^2$ up to a total derivative (there is an implicit sum over $i$ running 1 to $N$ on each derivative). As $D_i = \lambda_i \frac{\partial}{\partial \eta_i}$, the $D^2$ will only affect the $\eta$ pieces of $\mathcal O$, ignoring the harmonic structure of the $\lambda, \tilde \lambda$ tensor product explained above. The $D^2$ replaces two $\eta_i$ with $\lambda_i$. These two $\lambda_i$ and any free $\eta_i$ must combine into a totally antisymmetric $U(N)$ product in order to satisfy the supersymmetric Ward identity $Q A_N = 0$~\cite{Delgado:2023ogc}. This antisymmetric product is a column of height $N_\eta \ge 2$, a number that depends on the number of chiral superfields and derivatives. To form a legal YT, this new column must sit between the group of $N-2$ height columns and the height two columns.

\begin{figure}[h!]
\begin{center}
\includegraphics[scale=0.22]{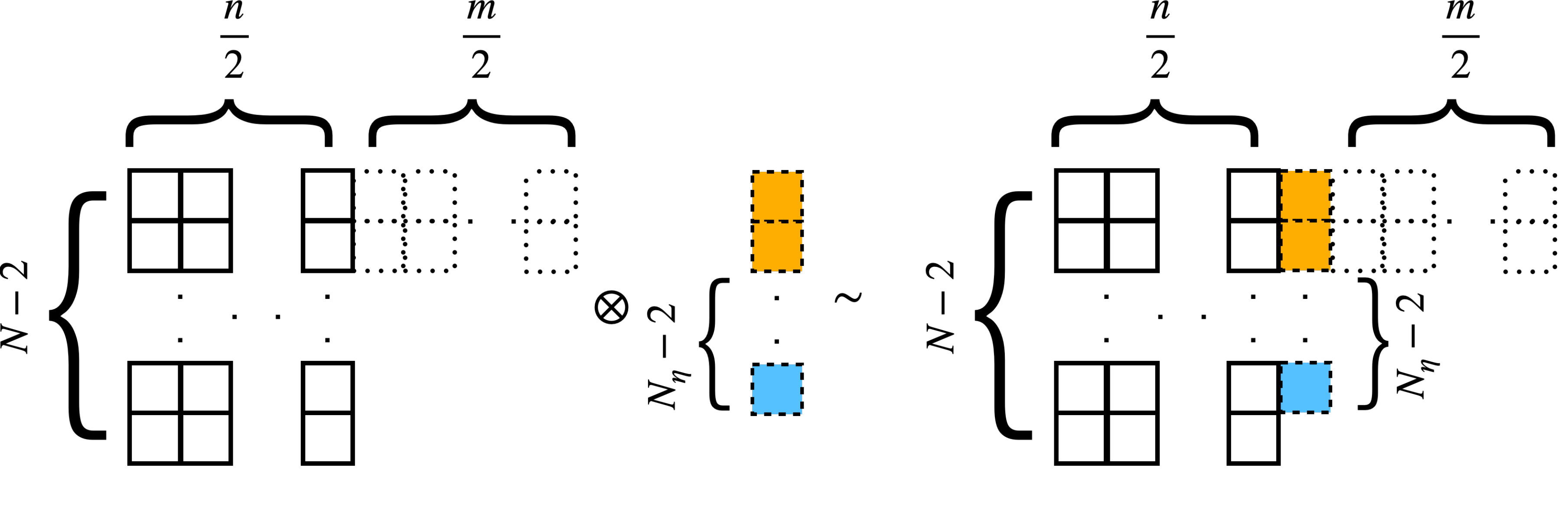}
\end{center}
\caption{The complete supersymmetric YT shape for $N_\eta \ge 2$ which satisfies $Q A_N = 0$.}
\label{sandwiching}
\end{figure}

 Astute readers may immediately notice that the above replacement automatically vanishes once there is no or only one $\eta$, i.e. $(\lambda_i \frac{\partial}{\partial \eta_i})^2(1)=\lambda_i \frac{\partial}{\partial \eta_i}(\eta_j)=0$. In this case one needs to go to the $\bar{\eta}$ basis either by performing a Grassmannian Fourier transformation or using the dual replacement rules under the $\bar{\eta}$ basis, which we discussed in our previous work \cite{Delgado:2023ogc}. In this paper we will only consider the operators containing more than one $\eta$ after applying replacement rules.

The final ingredient, $\overline D^2$ generates the delta function $\delta^2(Q^\dag)$ to enforce the $Q^\dag A_N = 0$ Ward identity. It is left off the diagram just as the total energy momentum delta function $\delta^4(P)$ is omitted from non-supersymmetric diagrams. The presence of the delta function is automatic for amplitudes in the $\eta$ basis, and Ref.~\cite{Delgado:2023ogc} showed explicitly how $\overline D^2$ leads to this factor in several examples.

To summarize -- we started with a selected operator from a class we were interested in, and the end result is a YT. For our representative operator, we know exactly what particle number indices accompany each spinor helicity variable, so we can even fill in the tableau with numbers.  To illustrate the steps outlined above consider the operator class $D^2\overline D^2 \Phi^2 (\Phi^\dag)^3$. Choosing the representative operator and replacing the fields with their spinor helicity form,
\begin{align}
D_\alpha \Phi_1 D^\alpha \Phi_2 \Phi^\dag_3 \overline D_{\dot\alpha} \Phi^\dag_4 \overline D^{\dot\alpha} \Phi^\dag_5 \to \lambda_{1\alpha}\lambda^\alpha_2\, \tilde \lambda_{4 \dot\alpha}\tilde \lambda^{\dot\alpha}_5\, \eta_4\, \eta_5 \to  (\epsilon_{12345} \tilde \lambda^4_{ \dot\alpha}\tilde \lambda^{5\dot\alpha })\, (\eta_4\, \eta_5)\, (\lambda_{1\alpha}\lambda^\alpha_2\,), 
\label{eq:SHex}
\end{align}
where we used $\epsilon_{12345}$ to convert all $U(N)$ indices to fundamentals in the last step. Each grouping in the line above corresponds to a YT column, and the groups have been rearranged so gluing their respective columns together gives a legal, harmonic YT. 
\begin{figure}[h!]
\begin{center}
\includegraphics[scale=0.28]{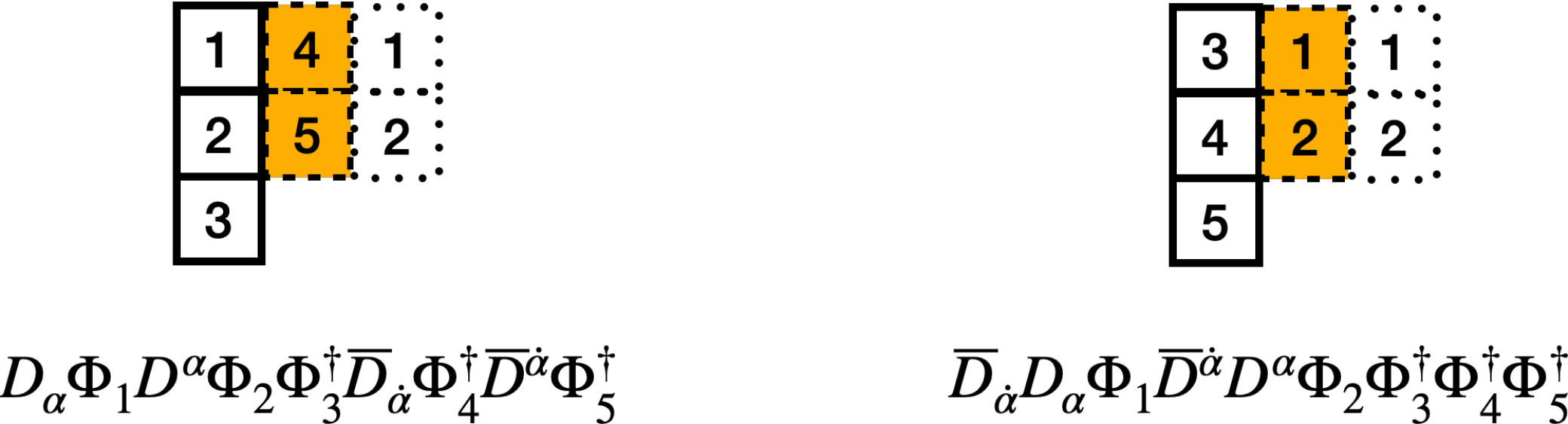}
\end{center}
\caption{Young tableaux for two different operators in the class $D^2\overline D^2 \Phi^2 (\Phi^\dag)^3$. The leftmost column contains the products of $\tilde \lambda_i$ (expressed in terms of $U(N)$ fundamentals using the $U(N)$ $\epsilon$ symbol), the right column contains the product of $\lambda_i$, and the middle, shaded column contains an additional product of $\lambda_i$ that originates from the $D^2 \in d^4\theta$. The indices in the boxes correspond to the $i$ index in e.g. $\lambda_i$ and change as we shuffle where the derivatives are placed.}
\label{egclass}
\end{figure}

Rather than listing the spinor contractions as in Eq.~\eqref{eq:SHex}, we can use angle and square bracket shorthand, $\langle 4 5 \rangle [12] [[45]]$, where we have introduced the notation $[[ij]]$ to indicate the product of $\eta$'s. Had we picked a different representative operator, say
\begin{align}
\label{eq:ex1b}
(\overline D_{\dot\alpha} D_\alpha \Phi_1)( \overline D^{\dot\alpha}D^\alpha \Phi_2) \Phi^\dag_3  \Phi^\dag_4  \Phi^\dag_5 \sim \langle 12 \rangle [12][[12]]
\end{align}
 we would have found a YT with the same shape but different filling.  The two different YT corresponding to Eq.~\eqref{eq:SHex} and \eqref{eq:ex1b} are shown below in Fig.~\ref{egclass}.

Now that we understand the process, we can skip the intermediate steps and go directly from the operator class to the YT shape. Consider an operator with $N$ chiral/antichiral superfields, $N_\Phi$ of which are chiral, along with $m$ $D$ and $n$ $\overline D$. Consulting the replacement rule, the numbers of $\lambda$ and $\tilde \lambda$ are clearly $m$ and $n$, implying $m/2$ and $n/2$ columns, while $N_\eta = N_\Phi - m + n$. With that information, and knowing the allowed YT shape, we can find the unique YT for the operator class. Including vector superfields, the counting changes to $(m+N_W)/2$  height 2 columns, $(n + N_{\bar W})/2$ height $N-2$ columns, and one column of height $N_\eta = N_\Phi + N_W - m + n$, where $N_W, N_{\bar W}$ are the number of $W$ and $\overline W$ superfields. 

Applied to the example above, $D^2\overline D^2 \Phi^2 (\Phi^\dag)^3$ has $n=m=2, N_W = N_{\overline W}=0, N = 5, N_\Phi=3$.  Using the counting above, this translates to a leftmost column of height $3$, a middle column of height $N_\eta = 2$ and a right hand column of height 2, as shown below in Fig.~\ref{yts_ex1}.
\begin{figure*}[h!]
\centering
\includegraphics[width=0.8\textwidth]{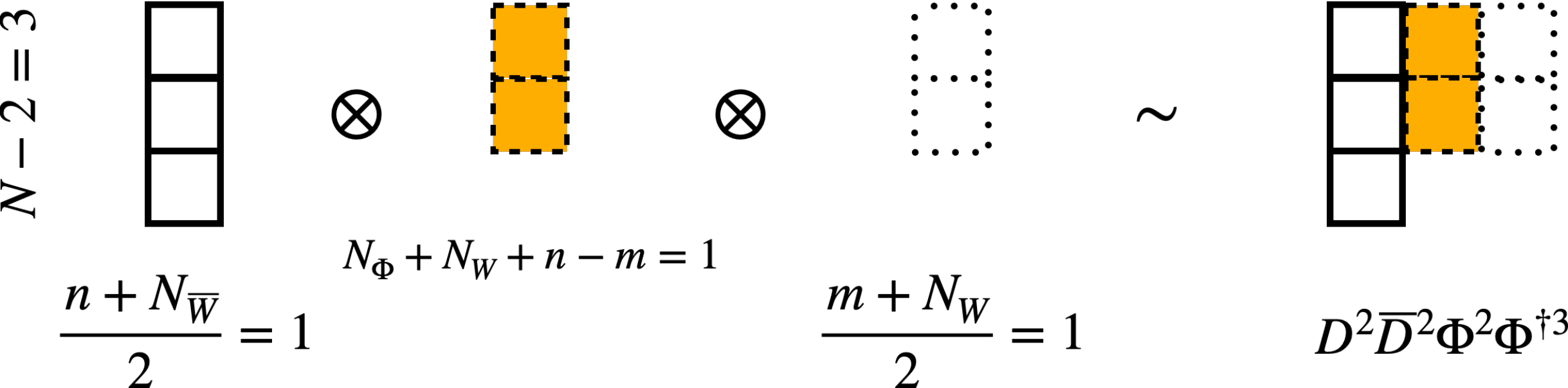}
\caption{Young tableau shape for  $D^2\overline D^2 \Phi^2 (\Phi^\dag)^3$ class operators. This shape is determined entirely by counting the number of chiral and anti-chiral superfields and the total number of derivatives and does not require us to pick a representative operator within the class to use as a guide. Notice that, at this stage we have no input into what index to put in each box.}
\label{yts_ex1}
\end{figure*}

\begin{figure*}[h!]
\centering
\includegraphics[width=0.8\textwidth]{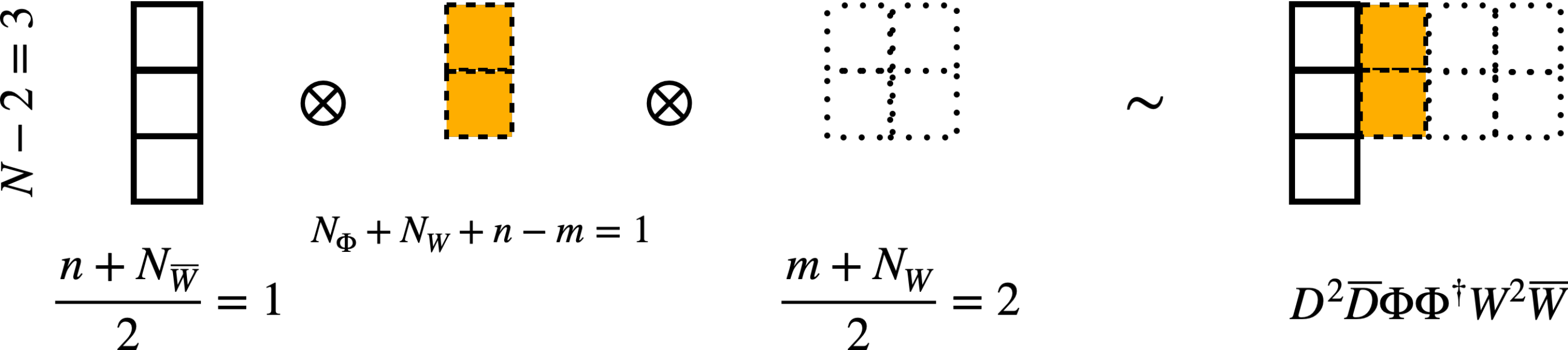}
\caption{Young tableau shape for $D^2\overline{D} \Phi\Phi^\dag W^2\overline{W}$ class operators. As in Fig.~\ref{yts_ex1}, the shape is determined purely by the number of fields of each type (chiral/vector) and the number of derivatives.}
\label{yts_ex2}
\end{figure*}

As a second example, consider $D^2\overline{D} \Phi\Phi^\dag W^2\overline{W}$. Here $m=2, n=1, N_W = 2, N_{\overline{W}}= 1, N=5, N_\Phi=1$, leading to a YT shape shown in Fig.~\ref{yts_ex2} shown below --  right hand column of height 3, followed by a column of height $N_\eta = 2$ representing the $\eta$ pieces, and a block of two height 2 columns.

We emphasize that the outcome of this class $\leftrightarrow$ YT translation is just the tableau shape. When we considered representative operators, these gave us the table shape and information on how the boxes should be filled with particle number indices. What we'd like is a way to go from the YT shape directly to a basis of operators for the class without having to think about all possible ways to partition derivatives and contract indices. Not only is the latter method tedious, it almost always over counts the number of independent operators as it is easy to miss redundancies due to group identities, integration by parts,  and the equations of motion. A more systematic method is presented below utilizing Semi-Standard Young tableaux.

\section{Semi-Standard Young Tableaux}\label{sec:ssytbasis}

Having reviewed how we go from an operator class to a unique YT shape, we move on to using the YT to find a basis for the operators in the corresponding class. By basis, we mean the set of independent operators that can be manipulated -- via integration by parts, the equations of motion, or group (Fierz/Schouten) identities -- into any operator with the same field and derivative content. 

The YT are indispensable for this task, as it is well known that an independent basis of a given shape and fillings of a Young tableau is formed with the following two rules:
\begin{itemize}
\item The numbers along a certain row weakly increase;
\item The numbers down a certain column strongly increase.
\end{itemize}
The result is called semi-standard Young tableaux (SSYT) basis~\cite{Frame1954TheHG, doi:10.1142/0097} and forms a natural basis of the $U(N)$ representations, i.e. operators in this paper. The number of SSYTs equals the dimension of each representation.\footnote{There are multiple bases for an YT, as one can always choose a linear combination among all elements.} In this paper we will only care about how to form the basis and we are not interested in finding a preferred basis, i.e. a basis under which certain calculations become easier. 

 The question then arises of how to find the fillings/numbers with a given superoperator/superamplitude without having to expand it in spinor helicity form. Explicitly, what we need are the number of entries, which we'll refer to as $\#i$ in the following, for each of the $N$ particles in the operator/amplitude (again, for now we are sticking to operators formed from distinguishable, massless fields).
 
  In non-supersymmetric (massless) theories, Ref~\cite{Li:2020gnx} showed that $\#i$ could be determined from the number of derivatives and the helicities of the particles involved:
\begin{align}
\#i = \frac{N_D}{2} + \sum\limits_{j,h_j >0}^N h_j - 2\, h_i 
\label{eq:numi}
\end{align}
where $N_D$ is the number of derivatives present, $h_j$ are the helicities ($0, \pm 1/2, \pm 1$) of the $N$ fields in the operator, and the sum extends only over the positive helicites. Once we know the complete set of numbers and fillings in the YT using these arguments, we immediately have the basis -- both its size and the explicit form of the operators -- just by knowing properties of the operator class. What we would like is the analog of Eq.~\eqref{eq:numi} for supersymmetric theories.

The algorithm above (Eq.~\eqref{eq:numi}) is, at its heart, just a mechanism for counting the number of $\lambda_i$ (or $\tilde \lambda^i$) without having to pick a representative operator from the class to expand as a guide. Note that counting the number of $\lambda_i$ (or $\tilde \lambda^i$) is really a proxy for counting the number of $U(N)$ fundamentals (antifundamentals) -- so when we look at superamplitudes we will want to count $\lambda_i$, $\tilde \lambda^i$ and $\eta_i$. Additionally (in both the supersymmetric and non-supersymmetric cases) we will use the $\epsilon_{1\cdots N}$ to relate any pair of $\tilde \lambda^i \tilde \lambda^i$ to $N-2$ fundamentals, so e.g for $N=5$, $\tilde \lambda^1 \tilde \lambda^2$ is equivalent, in terms of $U(N)$ indices, to $\lambda_3\lambda_4\lambda_5$.

Focusing first on operators composed purely of chiral/anti-chiral superfields --  operators classes ($N_\Phi$ chiral fields and $N_{\Phi^\dag}$ antichiral) --
\begin{align}\label{general form of super operator}
D^m\overline{D}^n \Phi_1\Phi_2\cdots\Phi_{N_\Phi} \Phi^\dag_{N_\Phi+1}\Phi^\dag_{N_\Phi+2}\cdots\Phi^\dag_{N_\Phi+N_{\Phi^\dag}},
\end{align}
let's examine how each piece, $D, \overline D, \Phi_i, \Phi^\dag_j$ contribute $U(N)$ indices. Reviewing the replacement rule, we see each $\Phi$ contributes one fundamental, while each $\Phi^\dag$ contributes nothing. Each $D$, removes one $\eta_i$ but replaces it with a $\lambda_i$, so no net change in the number of fundamentals. This leaves us with $\overline D = \tilde \lambda_i \eta_i$ as the only other factor contributing to index counting. Each pair of $\overline D$ (and for operators of the form above, $n$ must be even) generates $\tilde \lambda^i \tilde \lambda^j$, which, upon contracting with $\epsilon_{1\cdots N}$ generates a fundamental index for all of $N = N_\Phi +N_{\Phi^\dag}$ except for $i$ and $j$. However, the $\overline D$ pair also introduces $\eta_i \eta_j$, exactly compensating for the indices omitted when contracting the pair of $\tilde \lambda$ with $\epsilon$. Thus, we get an index for each chiral field plus an index for all (chiral and antichiral) fields for each pair of $\overline D$. In summary,
\begin{align}
\# i = \left\{ \begin{array}{cc} 1 + \frac{n}{2} & \text{for}\, \Phi_i \\ \frac{n}{2} & \text{for}\, \Phi^\dag_i \end{array}\right.. 
\label{eq:icount1}
\end{align}

Now let's allow vector superfields. The most general operator class we can consider is
\begin{align}\label{Standard form}
 D^m\overline{D}^n (\Phi)^{N_\Phi}(\Phi^{\dag})^{N_{\Phi^\dag}}(W)^{N_W} (\overline{W})^{N_{\overline W}}
\end{align}
where $n+N_{\overline W}$ and $m + N_W$ must be even by Lorentz invariance. Consulting Table~\ref{dicv}, each $W_i$ is straightforward to count as it introduces two fundamental indices. The $\overline W^i$ are more subtle, as they only introduce a single $\tilde \lambda^i$, and the procedure above assumed all $\tilde \lambda^i$ came in pairs and were each accompanied by an $\eta_i$. Had each $\overline W_i$ contained an $\eta_i$, the combination of $\overline D^n$ and $(\overline{W})^{N_{\overline W}}$ would look like (from spinor-helicity perspective), $n+N_{\overline W}$ derivatives, which we know from above gives $\frac{n+N_{\overline W}}{2}$ copies of all $N = N_\Phi+N_{\Phi^\dag}+N_W+N_{\overline W}$ indices. However, minus the accompanying $\eta_i$, we need to remove one index for each $\overline W_i$. Altogether,
\begin{align}
\# i = \left\{ \begin{array}{cc} 1 + \frac{n+N_{\overline W}}{2} & \text{for}\, \Phi_i \\
 \frac{n+N_{\overline W}}{2} & \text{for}\, \Phi^{\dag,i} \\
2 + \frac{n+N_{\overline W}}{2} & \text{for}\, W_i \\
\frac{n+N_{\overline W}}{2} -1 & \text{for}\, \overline W^i 
 \end{array}\right. .
 \label{eq:icount2}
\end{align} 
Note that the combination $(n+N_{\overline W})/2$ is the only thing that appears in $\#i$, while we need in addition the total number of fields $N$, the number $m$ of $D$ and the number of chiral superfields $N_\Phi + N_W$ to determine the YT shape.

\subsection{Examples}
Now that we have worked out how the YT shape and index counting ($\#1, 2, \cdots N$) depend on properties of the operator class, let us look a few examples.

For our first example, let us see operator class $D^2\overline D^2\Phi^2(\Phi^\dag)^3 \sim D^2\overline{D}^2\Phi_1\Phi_2\Phi_3^\dag\Phi_4^\dag\Phi_5^\dag$ through to the end. We need the latter expression to know which particle number indices belong to chiral fields and which belong to antichiral fields. The YT shape for this operator class has been shown in Fig.~\ref{yts_ex1}, derived from  $N = 5, m = n = 2, N_\Phi =2$ and $N_{\Phi^\dag} = N-N_\Phi = 3$. Following Eq.~\eqref{eq:icount1}, these seven boxes should be filled with indices $i$, where $\#i = 2$ for each chiral superfield and $\#i = 1$ for each anti-chiral, so $\{1,1,2,2,3,4,5\}$. There are six possible SSYT fillings, shown below in Fig.~\ref{ex1}.
\begin{figure}[h!]
\begin{center}
\includegraphics[scale=0.28]{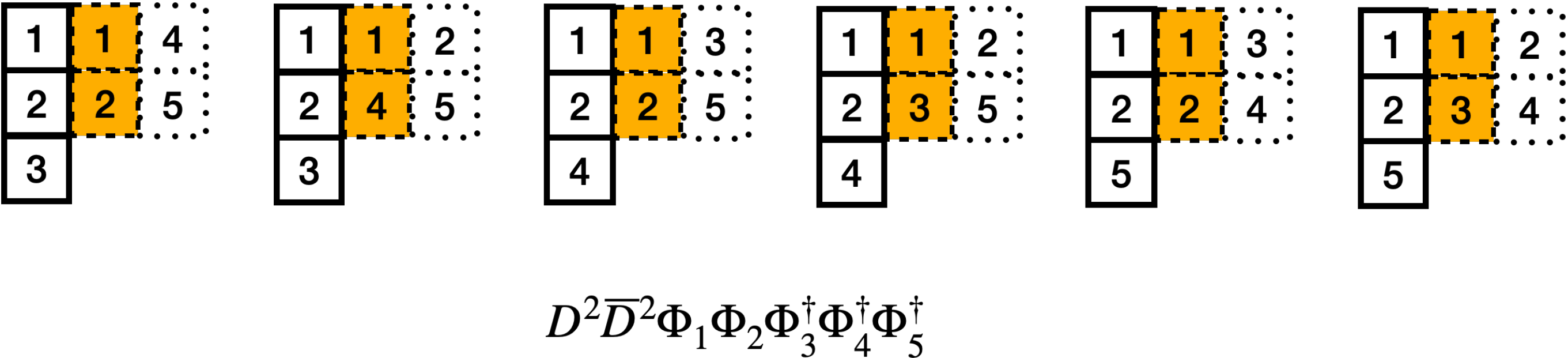}
\end{center}
\caption{The SSYT basis for the operator class $\sim D^2\overline{D}^2\Phi_1\Phi_2\Phi_3^\dag\Phi_4^\dag\Phi_5^\dag$. The spinor helicity form for each independent operator can be read directly from the diagram, and can be converted to superfield format as explained in the text and in Ref.~\cite{Delgado:2023ogc}. } 
\label{ex1}
\end{figure}
The number of independent operators -- $6$ -- matches what we get using the Hilbert series method~\cite{Delgado:2022bho, Delgado:2023ivp}. However, using the SSYT approach, we also get the form of each basis operator.  The spinor helicity form can be read off directly, e.g. $\langle45\rangle[45][[12]]$ for the leftmost operator in Fig.~\ref{ex1}, where we use [[$\cdots$]] to represent the 'supersymmetrization' piece of the diagram. This can be converted to superoperator form by dropping the [[$\cdots$]] and mapping $|i ] \to D_i, |i \rangle \to \overline D_i$, e.g. $\langle45\rangle[45][[12]] \rightarrow \Phi_1\Phi_2\Phi_3^\dag D\overline{D}\Phi_4^\dag D\overline{D}\Phi_5^\dag$. See Ref.~\cite{Delgado:2023ogc} for more details on the spinor helicity to superoperator tranlastion, and in particular how to adapt the procedure when vector superfields are present.  

Note that, while the particular form of the filled YT -- the \#i -- depends on our choice to label the fields as $\Phi_1\Phi_2\Phi_3^\dag\Phi_4^\dag\Phi_5^\dag$, beginning with the chiral superfields sequentially starting with index \#1, followed by the anti-chiral fields, the counting is independent of this choice. We would get six operators applying the SSYT to any other labeling choice, such as $\sim D^2\overline{D}^2\Phi_4\Phi_5\Phi_1^\dag\Phi_2^\dag\Phi_3^\dag$ or $\sim D^2\overline{D}^2\Phi_1\Phi_5\Phi_3^\dag\Phi_2^\dag\Phi_4^\dag$.\footnote{Said another way,  by construction, the counting of SSYT stays the same under an arbitrary permutation of indices/labels.} A different ordering does result in a different basis, and it may be the case that one choice is better than another for certain calculations, however for our purposes we only care about finding ``a" basis, so we are free to pre-order the labels as in \eqref{general form of super operator}

As a second example, we add an additional chiral superfield -- $D^2\overline{D}^2\Phi_1\Phi_2\Phi_3\Phi_4^\dag\Phi_5^\dag\Phi_6^\dag$, so $N = 6, m = n = 2, N_\Phi = 3 = N_\eta$ and $N_{\Phi^\dag} = N-N_\Phi = 3$. The YT now has 9 boxes, arranged into a column of height $N-2 = 4$ from $\overline D$, a column of height $N_\eta = 3$ and a column of height 2 from the $D$. The $\#i$ counting is identical to the first example, there are just more chiral fields -- $\#i = \{1,1,2,2,3,3,4,5,6\}$. Filling the YT, we find 17 SSYT, corresponding to a basis of 17 operators. The fillings are shown below in Fig.~\ref{ex2}

\begin{figure}[h!]
\begin{center}
\includegraphics[scale=0.3]{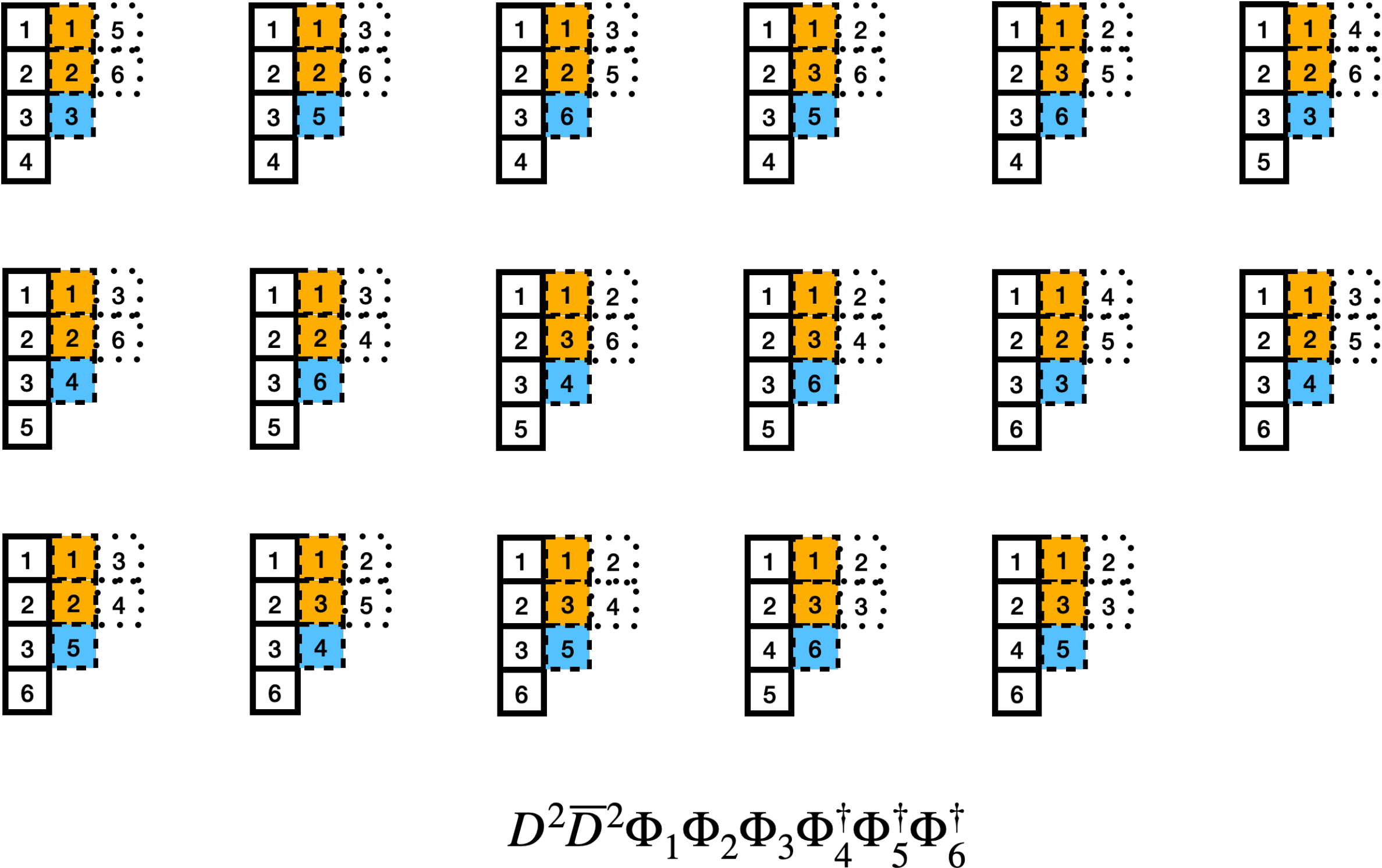}
\end{center}
\caption{The SSYT basis, 17 total operators, for the operator class $\sim D^2\overline{D}^2\Phi_1\Phi_2\Phi_3\Phi_4^\dag\Phi_5^\dag$. As in the previous example, while the operator form (either as YT or superfields) will depend on how we label the fields, meaning which we take as field \#1, which as \#2, etc., the number of independent operators does not depend on this choice.}
\label{ex2}
\end{figure}

Again one can directly read the operator SH form directly from the YT.  For example, the first diagram reads: $\langle56\rangle[56][[123]]\equiv\langle56\rangle[56]\sum_{perm=1,2,3}([[ij]]\eta_k)$, which corresponds to the operator $\Phi_1\Phi_2\Phi_3\Phi_4^\dag D\overline{D}\Phi_5^\dag D\overline{D}\Phi_6^\dag$.

For our final example, we return to $D^2\overline{D}^{\dot{\alpha}} \Phi_{1}\Phi^\dag_2W_{3}^{\alpha}W_{4\alpha}\overline{W}_{5\dot{\alpha}}$. The YT shape for this class can be found in Fig.~\ref{yts_ex2}, using  $N=5, m=2, n =1$, $N_\Phi = N_{\Phi^\dag} = N_{\overline W} = 1$ and $N_W = 2$. These 9 boxes are to be filled with $\#i$ following Eq.~\eqref{eq:icount2} with $(n+N_{\overline W})/2 = 1$: $\#i = \{1,1,2,3,3,3,4,4,4\}$. The three possible SSYT fillings are shown below in Fig.~\ref{ssytvector}, and correspond to the operators $\langle2 5 \rangle[24][34][[13]] = \Phi_{1}D^\beta\overline{D}^{\dot{\alpha}}\Phi^\dag_2W_{3}^{\alpha}D_\beta W_{4\alpha}\overline{W}_{5\dot{\alpha}}$, $\langle 4 5 \rangle [34]^2[[14]] = \Phi_{1}\Phi^\dag_2 D_\beta W_{3}^{\alpha}\overline{D}^{\dot{\alpha}}D^\beta W_{4\alpha}\overline{W}_{5\dot{\alpha}}$, and $\langle 3 5 \rangle [34]^2[[13]] = \Phi_{1}\Phi^\dag_2\overline{D}^{\dot{\alpha}}D^\beta W_{3}^{\alpha}D_\beta W_{4\alpha}\overline{W}_{5\dot{\alpha}}$.

\begin{figure}[h!]
\begin{center}
\includegraphics[scale=0.27]{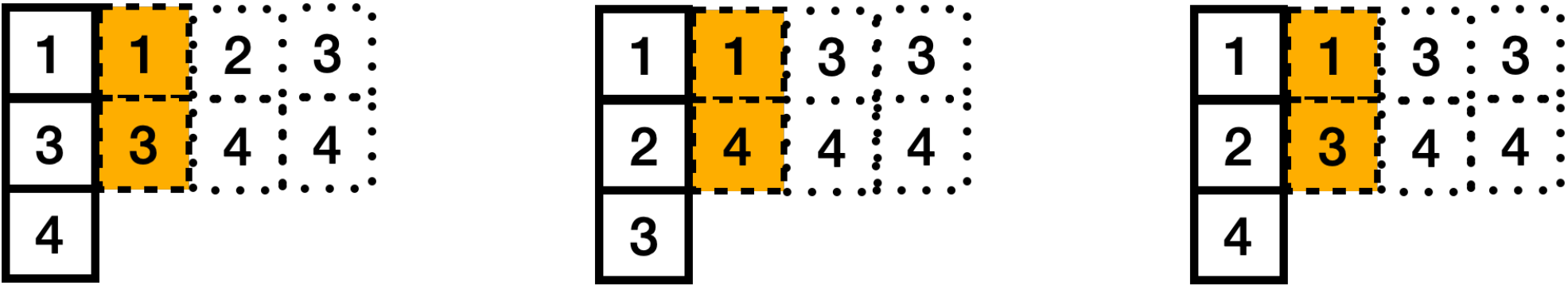}
\end{center}
\caption{The SSYT basis, 3 total operators, for the operator class $D^2\overline{D}^{\dot{\alpha}} \Phi_{1}\Phi^\dag_2W_{3}^{\alpha}W_{4\alpha}\overline{W}_{5\dot{\alpha}}$.}
\label{ssytvector}
\end{figure}

Notice that in previous studies \cite{Henning:2019mcv,Henning:2019enq} one reads off the amplitude/operator from a given diagram by taking certain symmetrization/antisymmetrization among columns and rows, resulting in an amplitude which is annihilated by the conformal generator $K=\frac{\partial}{\partial\lambda}\frac{\partial}{\partial\tilde{\lambda}}$. However one can always use total (super-)momentum conservation to add arbitrary polynomials of momentum $P=\lambda\tilde{\lambda}$ and leave the on-shell expression unchanged. Therefore the easiest choice is simply to take the fillings and form a monomial following the rules~\cite{Li:2022tec}. This is the 'trick' we apply when reading the diagrams.

\section{Indistinguishable Superfields}\label{Indistinguishable Superfields}

Having discussed the construction of a basis for higher dimensional supersymmetric operators formed from (massless), distinguishable superfields, we now turn to the case where identical superfields appear in the operator. 

Formally, one can pick the basis for operators with identical superfields by taking the quotient of operator basis with distinguishable superfields with respect to the following equivalence relation(s):
\begin{equation}\label{equivlence relation}
\mathcal{O}(\Phi_{1,2,\cdots},\Phi^\dag_{1,2,\cdots})\sim \mathcal{O}'(\Phi_{1,2,\cdots,m-1,m=n,m+1,\cdots},\Phi^\dag_{1,2,\cdots,m-1,m=n,m+1,\cdots}),
\end{equation}
assuming $\Phi_{m}(\Phi^\dag_{m})$ is identified as another superfield $\Phi_{n}(\Phi^\dag_{n})$, where $\mathcal{O}'$ denotes the operator with indistinguishable superfields. One operator $\mathcal{O}$ may be identified with several $\mathcal{O}'$s\footnote{In terms of spinor helicity variables, this relation leads to a linear relation with two or more terms. A concrete example is given later in Eq.\eqref{an} to \eqref{coe1}.}, which indicates that not all of them are linearly independent, and we need to remove such redundancies when constructing the basis.

It is straightforward to see that the existence of indistinguishable superfields will lead to fewer independent superamplitudes, since the new symmetry group $\mathcal{G}$ now is not the entire $U(N)$ symmetry among $\lambda,\tilde{\lambda},\eta$, but becomes the quotient space because of the $S_{n_i}$ groups among identical fields, with $S_{n_i}$ being the usual symmetric group and $n_i$ is the multiplicity of each field, 
\begin{equation}
\mathcal{G}=U(N) / \prod_i^N S_{n_i}.
\end{equation}
To construct a basis with symmetry $\mathcal{G}$, the recipe we follow is: construct the basis with all distinguishable superfields and then breaks $U(N)\rightarrow \mathcal{G}$, which is the actual symmetry when identical superfields are involved. The $S_n$ symmetry is removed by calculating the quotient space with the equivalence relations defined in Eq. (\ref{equivlence relation}).

One may wonder if we could directly construct the SSYTs for $\mathcal{G}$ instead of the single $U(N)$. After all, Young diagrams can describe $S_n$ symmetry and we should expect that $\mathcal{G}$ as a quotient shares a similar property. However, this topic is beyond the scope of the current paper.\footnote{Garnir relations \cite{Fulton1996} may be helpful and give some insight for readers who are interested in this direction.} For now, we will apply a somewhat brute force matrix approach to deal with indistinguishable superfields. Despite being cumbersome, this approach is general and should work for any case.\footnote{For cases involving more global/gauge symmetries, one may refer to \cite{Li:2020gnx} where a rigorous tensor representation approach is built for non-supersymmetric EFTs and it's not difficult to generalize to the supersymmetric we study here.}

\subsection{General Approach}
For a given operator class containing $N$ fields, we first form the SSYT basis following the steps in Sec.~\ref{results},\, \ref{sec:ssytbasis} assuming all particles are distinguishable. Each particle (or state, when the operators are viewed as a non-factorizable amplitudes) comes with its own label, which we use to fill the YT following the SSYT prescription. We then combine the basis elements into a vector  $\vec{A}=(A_1,A_2,\cdots,A_n)^T$, where $A_i=A_i(\lambda,\tilde{\lambda},\eta)$ corresponds to the $i$th YT/amplitude and $n$ is the basis size for the operator class in question. From this setup, we want to make two states -- which formerly carried labels $a$ and $b \in 1\cdots N$ --  identical.

Starting with an amplitude of distinguishable states, let's act on it with $S_{2}$, interchanging the labels of the states ($a \leftrightarrow b$). The resulting amplitude will not necessarily be an element of the SSYT basis $\vec A$, but as the basis is complete we can always express  it as a linear combination of the $A_i$. Mathematically, we phrase this as $\vec A  \xrightarrow[S_2]{} H\vec{A}$, such that the $i^{th}$ row of $H$ gives the linear combination of basis operators that $A_i$ transforms into under $S_2$. 

However, if two states are identical, how we label them in an amplitude should not matter, meaning swapping what we call state $a$ and state $b$ must get us the same amplitude up to a sign -- $+1$ for bosonic states and $-1$ for fermionic. This statement is true regardless of the number of states present or the nature of the interactions (meaning whether they are momentum-dependent). In terms of the $\vec A$, we can express this as $\vec{A}\rightarrow M\vec{A}$ under label exchange, where $M =diag(m_{jj})$ and $m_{jj}=\pm1$. 

Demanding 
\begin{align}
H\vec A = M\vec A, \quad\text{or}\quad (H-M)\vec A = 0
\label{eq:indis}
\end{align}
sets the amplitudes with labels $a \leftrightarrow b$ exchanged equal to $\pm$ the original amplitudes -- the result when $a$ and $b$ represent indistinguishable states. The matrix equation we get from imposing \eqref{eq:indis} tells us what relations among the $A_i$ result from $a \leftrightarrow b$ indistinguishability. First off, $H-M$ may have rows that are all $0$. These tell us nothing as Eq.~\eqref{eq:indis} is satisfied for all $\vec A$, so let us remove them by row reducing. This gets us a $p \times n$ matrix we call $T$, where $p = rank(H-M)$ and $n$ is the number of amplitudes in $\vec A$. As $T \ne 0$, Eq.~\eqref{eq:indis} can only be satisfied if there are relations among the $A_i$. These relations are collected in the null-space vectors $P$ of $T$.\footnote{A nullspace vector defines a linear combination of operators which vanishes under IBP and EOM equivalence relations.}

The procedure we just described works for two identical particles. If three or more idential particles, the symmetry group is $S_{n>2}$, we then break $S_{n>2}$ into generators (labeled by $k_i$) and repeat the procedure leading up to Eq.\eqref{eq:indis} for each generator: $(H_{k_i} - M_{k_i})\vec A = 0$. For each $k_i$, we can row reduce and find the non-zero matrix $T_{k_i}$, then stack them together into a single $(\sum_{k_i} rank(T_{k_i})) \times n$ matrix. The null-space of this combined  $T$ tells us the relations among operators under the full $S_{n>2}$.

The $M$ matrix merits further comment because it depends on the type of superfield. For identical anti-chiral superfields $M = +\mathbf 1_{n\times n}$ (meaning each amplitude goes to itself under interchange of identical particles), while for chiral superfields $M = -\mathbf 1_{n\times n}$. Why are the signs in $M$ opposite for chiral superfield and anti-chiral superfield if they are both 'scalar' fields? We should remember that we are working under $\eta$-representation and chiral superfields under this parametrization are associated with an inherent Grassmann variable. The upshot here is that assigning $\eta$'s to chiral superfields makes them Grassmann and we cannot naively exchange the two fields without adding a minus sign. In the case of anti-chiral superfields, this is not a problem because they are still 'numbers'.\footnote{If we work under $\bar{\eta}$-representation, exchanging two anti-chiral superfields will introduce a minus sign while chiral superfields are treated as 'numbers'.}  To better illustrate the procedure, let us work through some examples

\subsection{Examples}
We will study the case $D^2\overline{D}^2\Phi_1\Phi_2\Phi_3^\dag\Phi_4^\dag\Phi_5^\dag$ for example. Recall the six independent terms we got in the previous section:
\begin{equation}
\begin{split}
&\Phi_1\Phi_2\Phi_3^\dag D\overline{D}\Phi_4^\dag D\overline{D}\Phi_5^\dag,\quad \Phi_1D\Phi_2\Phi_3^\dag \overline{D}\Phi_4^\dag D\overline{D}\Phi_5^\dag,\quad
\Phi_1\Phi_2D\overline{D}\Phi_3^\dag \Phi_4^\dag D\overline{D}\Phi_5^\dag,\\
&\Phi_1D\Phi_2\overline{D}\Phi_3^\dag \Phi_4^\dag D\overline{D}\Phi_5^\dag,\quad\Phi_1\Phi_2D\overline{D}\Phi_3^\dag D\overline{D}\Phi_4^\dag \Phi_5^\dag,\quad\Phi_1D\Phi_2\overline{D}\Phi_3^\dag D\overline{D}\Phi_4^\dag \Phi_5^\dag.
\end{split}
\end{equation}
which we can write in terms of $\lambda,\tilde{\lambda},\eta$ and label $A_1 \cdots A_6$:
\begin{equation}\label{an}
\begin{split}
A_1 = &[45]\langle45\rangle[[12]],\quad A_2 = [25]\langle45\rangle[[14]],\quad A_3 = [35]\langle35\rangle[[12]],\quad \\
A_4 = &[25]\langle35\rangle[[13]],\quad A_5 =  [34]\langle34\rangle[[12]],\quad A_6 = [24]\langle34\rangle[[13]].
\end{split}
\end{equation}

Let us consider the case  where $\Phi_1$ and $\Phi_2$ are identical chiral superfields, which means a $S_2$ symmetry. To determine the matrix $H$, the first step is to act on $A_i$ with the $S_2$ permutation (generator) $(12)$ (which swaps $1 \leftrightarrow 2$), labelling the result $B_i$:
\begin{equation}\label{bn1}
\begin{split}
B_1 = &[45]\langle45\rangle[[21]],\quad B_2 = [15]\langle45\rangle[[24]],\quad B_3 = [35]\langle35\rangle[[21]],\quad \\
B_4 = &[15]\langle35\rangle[[23]],\quad B_5 = [34]\langle34\rangle[[21]],\quad B_6 = [14]\langle34\rangle[[23]].
\end{split}
\end{equation}
Using properties of spinor products such as antisymmetry and the Schouten identity, the $B_i$ can be re-expressed as combinations of the $A_i$:
\begin{equation}\label{coe1}
\begin{split}
&B_1=-A_1,\ \ B_2=A_2-A_1\\
&B_3=-A_3,\ \ B_4=A_4-A_3\\
&B_5=-A_5,\ \ B_6=A_6-A_5.
\end{split}
\end{equation}
This action under $(12)$ can be expressed in matrix form as $\vec{A} \xrightarrow[S_2]{} \vec B = H\vec A$ where:
\begin{align}
H=
\begin{pmatrix}
-1 & \ 0 & 0 & \ 0 & 0 & \ 0\\
-1 & \ 1 & 0 & \ 0 & 0 & \ 0\\
0 & \ 0 & -1 & \ 0 & 0 & \ 0\\
0 & \ 0 & -1 & \ 1 & 0 & \ 0\\
0 & \ 0 & 0 & \ 0 & -1 & \ 0\\
0 & \ 0 & 0 & \ 0 & -1 & \ 1
\end{pmatrix}.\ \ 
\end{align}
Next, since $\Phi_1$ and $\Phi_2$ are chiral superfields, $\vec A \rightarrow M \vec A,$ where $M = -\mathbf 1_{6\times 6}$. From these two matrices, we can form $F=H-M$ and row reduce to get rid of the zero rows, leaving $T$.

\begin{equation}
F=
\begin{pmatrix}
0 & \ 0 & 0 & \ 0 & 0 & \ 0\\
-1 & \ 2 & 0 & \ 0 & 0 & \ 0\\
0 & \ 0 & 0 & \ 0 & 0 & \ 0\\
0 & \ 0 & -1 & \ 2 & 0 & \ 0\\
0 & \ 0 & 0 & \ 0 & 0 & \ 0\\
0 & \ 0 & 0 & \ 0 & -1 & \ 2
\end{pmatrix},\ \ 
T=
\begin{pmatrix}
-1 & \ 2 & 0 & \ 0 & 0 & \ 0\\
0 & \ 0 & -1 & \ 2 & 0 & \ 0\\
0 & \ 0 & 0 & \ 0 & -1 & \ 2
\end{pmatrix}.
\end{equation}
The nullspace vectors for $T$ are given by:
\begin{equation}
p_1=(2,1,0,0,0,0)^T,p_2=(0,0,2,1,0,0)^T,p_3=(0,0,0,0,2,1)^T.
\end{equation}
From the three nullspace vectors $p_{1,2,3}$ we get the relations
\begin{equation}\label{equiva}
A_1\sim2A_2\ \ , A_3\sim2A_4\ \ , A_5\sim2A_6.
\end{equation}
Each '$\sim$' defines an equivalence relation between two operators (with respect to IBP and EOM) after the identification of the two chiral superfields. We get three equivalence relations from \eqref{equiva}, so the number of independent operators reduces from six to three. The three independent basis vector therefore are easily read off as $\alpha_1 A_1+\alpha_2 A_2$, $\alpha_3A_3+\alpha_4A_4$, $\alpha_5A_5+\alpha_6A_6$, or any non-trivial linear combination of the three, provided $\alpha_{1,3,5}\neq-2\alpha_{2,4,6}$ respectively. 

As a second example, let us work out the procedure when more than two superfields are identical. The example we will study is still the $D^2\overline{D}^2\Phi_1\Phi_2\Phi_3^\dag\Phi_4^\dag\Phi_5^\dag$, but now we take the indistinguishable superfields to be $\Phi_3^\dag,\Phi_4^\dag,\Phi_5^\dag$. In this case the symmetry group is $S_3$ and we take the generators to be $(34)$ and $(45)$. Let us first act $(34)$ on Eq.~\eqref{an}, yielding
\begin{equation}\label{bn2}
\begin{split}
&[35]\langle35\rangle[[12]],\quad [25]\langle35\rangle[[13]],\quad [45]\langle45\rangle[[12]],\quad \\
&[25]\langle45\rangle[[14]],\quad [43]\langle43\rangle[[12]],\quad [23]\langle43\rangle[[14]].
\end{split}
\end{equation}
Following the previous procedure, we can find the matrices $N_{(34)}$, $M_{(34)}$ and the nullspace $T_{(34)}$:
\begin{equation}
H_{(34)}=
\begin{pmatrix}
\ 0 & \ 0 & \ 1 & \ 0 & \ 0 & \ 0\\
\ 0 & \ 0 & \ 0 & \ 1 & \ 0 & \ 0\\
\ 1 & \ 0 & \ 0 & \ 0 & \ 0 & \ 0\\
\ 0 & \ 1 & \ 0 & \ 0 & \ 0 & \ 0\\
\ 0 & \ 0 & \ 0 & \ 0 & \ 1 & \ 0\\
\ 0 & \ 0 & \ 0 & \ 0 & \ 1 & -1
\end{pmatrix},\ \ 
M_{(34)}=+\mathbf 1_{6\times6},\ \ T_{(34)}=
\begin{pmatrix}
\ 1 & \ 0 & \ -1 & \ 0 & \ 0 & \ 0\\
\ 0 & 1 & \ 0 & \ -1 & \ 0 & \ 0\\
\ 0 & \ 0 & \ 0 & \ 0 & \ 1 & \ -2
\end{pmatrix}
\end{equation}

We then repeat the procedure for the generator $(45)$:
\begin{equation}
H_{(45)}=
\begin{pmatrix}
\ 1 & \ 0 & \ 0 & \ 0 &\  0 & \ 0\\
\ 1 &  -1 & \ 0 & \ 0 & \ 0 & \ 0\\
\ 0 & \ 0 & \ 0 & \ 0 & \ 1 & \ 0\\
\ 0 & \ 0 & \ 0 & \ 0 & \ 0 & \ 1\\
\ 0 & \ 0 & \ 1 & \ 0 & \ 0 & \ 0\\
\ 0 & \ 0 & \ 0 & \ 1 & \ 0 & \ 0
\end{pmatrix},\ \ 
M_{(45)}=+\mathbf 1_{6\times6},\ \ T_{(45)}=
\begin{pmatrix}
\ 1 & -2 & \ 0 & \ 0 & \ 0 & \ 0\\
\ 0 & 0 & \ 1 & \ 0 &  -1 & \ 0\\
\ 0 & \ 0 & \ 0 & \ 1 & \ 0 & \ -1
\end{pmatrix}
\end{equation}

We then combine $T_{(34)}$ and $T_{(45)}$ together to form $T$: 
\begin{equation}
T=
\begin{pmatrix}
\ 1 & \ 0 & \ -1 & \ 0 & \ 0 & \ 0\\
\ 0 & 1 & \ 0 & \ -1 & \ 0 & \ 0\\
\ 0 & \ 0 & \ 0 & \ 0 & \ 1 & \ -2\\
\ 1 & -2 & \ 0 & \ 0 & \ 0 & \ 0\\
\ 0 & 0 & \ 1 & \ 0 &  -1 & \ 0\\
\ 0 & \ 0 & \ 0 & \ 1 & \ 0 & \ -1
\end{pmatrix},
\end{equation}
which has nullspace $P=(2,1,2,1,2,1)^T$. Same as before one gets the following relations among the six operators: $A_1\sim A_3 \sim A_5\sim 2A_2\sim 2A_4\sim 2A_6$, which reduces the number of independent operators from 6 to 1. Therefore we know that any operator in the original basis can be selected as the basis when antichiral superfields are identified.

\section{Conclusion and Discussion}\label{conclusions}

\begin{figure}[h!]
\begin{center}
\includegraphics[scale=0.27]{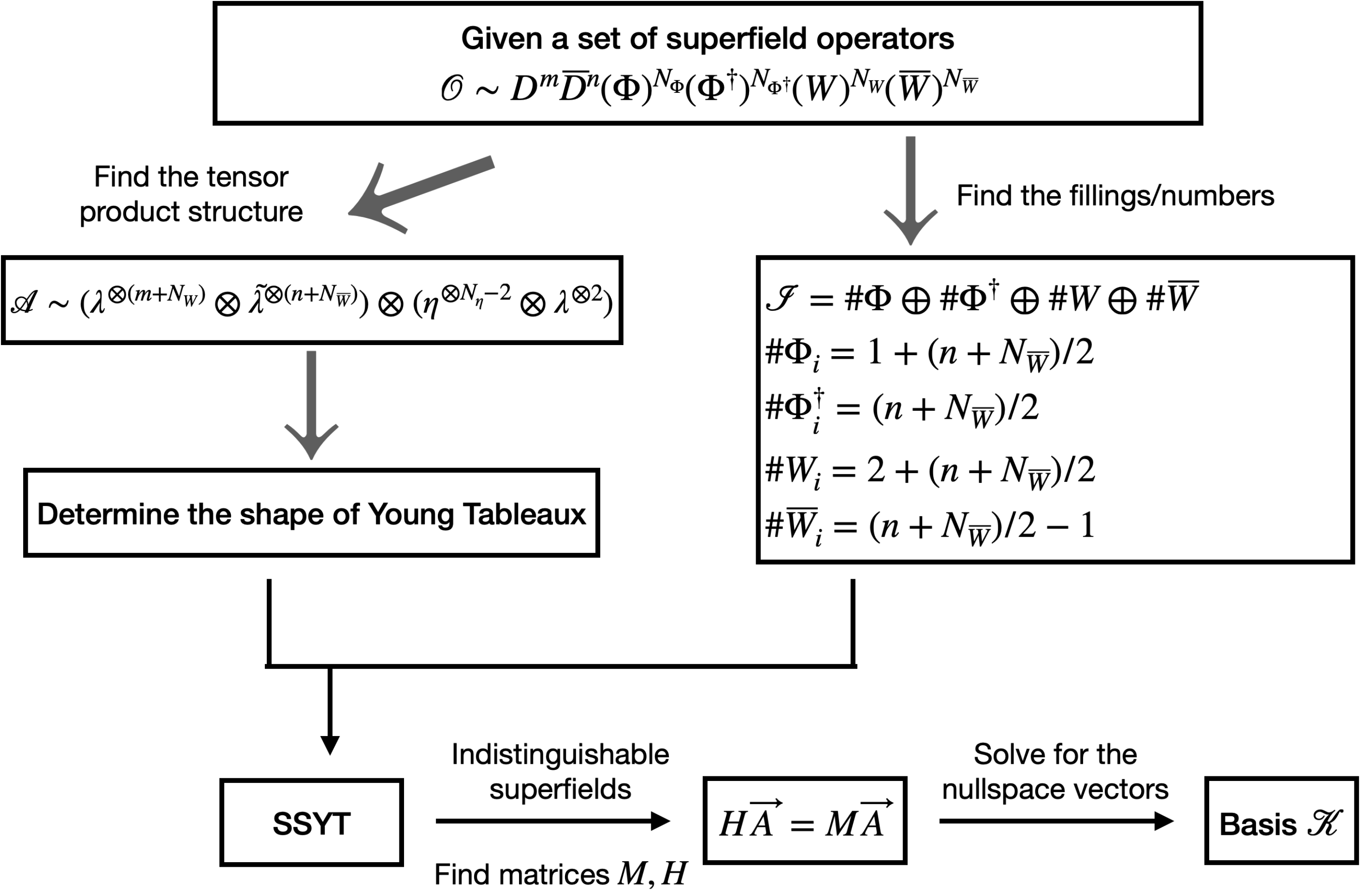}
\end{center}
\caption{Flowchart summarizing the results of our paper.}
\label{ssytflowchart}
\end{figure}

In this paper we have developed a SSYT approach to construct a basis of $\mathcal{N}=1$ supersymmetric effective operators  at arbitrary mass dimensions with any number of distinguishable/indistinguishable superfields and superderivatives. We have proven the relation between classes of operators and SSYT using an internal $U(N)$ symmetry ($N$ = number of fields in the operator) that acts on helicity amplitude variables and supersymmetric state Grasmmanian variables.  Given the contents of the target operator space, we first treat all superfields as distinguishable and order them in the standard form (see eq.~\ref{Standard form}). The number of undotted and dotted indices, combined with the number of chiral superfields determine the shape of YT, while the number of chiral superfields and derivatives determine the supersymmetric part of the complete YT under $\eta$-representation (check Fig. \ref{sandwiching} and the text around). The power of each building blocks, set by the number of $\overline{D}$ and $\overline{W}$ determines the set of numbers to fill the boxes, i.e. eq. \eqref{eq:icount1} and \eqref{eq:icount2}. Finding all SSYTs is straightforward and these diagrams correspond to a basis free of EOM and IBP redundancies. If  there are identical superfields, we can pick the subspace spanned by the set of identical superfields and remove the redundant operators using the matrix approach introduced in Section. \ref{Indistinguishable Superfields}. We summarize the steps in the flowchart Fig. \ref{ssytflowchart}.

One can already notice that the SSYT approach becomes complicated when indistinguishable superfields are involved, and this clearly comes from the fact that one cannot construct a diagram without labelling fields, but the labelling explicitly breaks the permutation symmetry. 

If one only wants to find the dimension of the basis for a given supersymmetric operator class, a recent approach via the Hilbert series \cite{Delgado:2022bho,Delgado:2023ivp} is more effective. Whether or not the superfields are indistinguishable doesn't complicate the calculation, in contrast to the SSYT approach we studied here, where we need to find matrices $H,M$ for indistinguishable cases. However Hilbert series method has its own problem: the calculations become unwieldy if the number of derivatives or fields of the operator class becomes large. The two approaches give exactly the same counting of independent operators at any given mass dimensions, and serve as complementary methods and cross-checks.

Some future avenues of research along this direction are: extending this approach to the massive case, where the little group for each state in an amplitude is $SU(2)$ rather than $U(1)$; studying the possible recursive pattern between amplitudes; relating component amplitudes and supersymmetrization, or analyzing the possible role of superconformal or dual superconformal~\cite{Drummond:2008vq} symmetry.

\acknowledgments

This is partially supported by the National Science Foundation under Grant Number PHY-2112540.

\appendix

\bibliographystyle{utphys}
\bibliography{ref}

\providecommand{\href}[2]{#2}\begingroup\raggedright\begin{thebibliography}{10}

\bibitem{Delgado:2023ogc}
A.~Delgado, A.~Martin, and R.~Wang, ``{Hidden U(N) symmetry behind $
  \mathcal{N} $ = 1 superamplitudes},''
  \href{http://dx.doi.org/10.1007/JHEP11(2023)215}{{\em JHEP} {\bf 11} (2023)
  215}, \href{http://arxiv.org/abs/2309.15802}{{\tt arXiv:2309.15802
  [hep-th]}}.

\bibitem{Delgado:2022bho}
A.~Delgado, A.~Martin, and R.~Wang, ``{Constructing operator basis in
  supersymmetry: a Hilbert series approach},''
  \href{http://dx.doi.org/10.1007/JHEP04(2023)097}{{\em JHEP} {\bf 04} (2023)
  097}, \href{http://arxiv.org/abs/2212.02551}{{\tt arXiv:2212.02551
  [hep-th]}}.

\bibitem{Delgado:2023ivp}
A.~Delgado, A.~Martin, and R.~Wang, ``{Counting operators in N = 1
  supersymmetric gauge theories},''
  \href{http://dx.doi.org/10.1007/JHEP07(2023)081}{{\em JHEP} {\bf 07} (2023)
  081}, \href{http://arxiv.org/abs/2305.01736}{{\tt arXiv:2305.01736
  [hep-th]}}.

\bibitem{Lehman:2015via}
L.~Lehman and A.~Martin, ``{Hilbert Series for Constructing Lagrangians:
  expanding the phenomenologist's toolbox},''
  \href{http://dx.doi.org/10.1103/PhysRevD.91.105014}{{\em Phys. Rev. D} {\bf
  91} (2015)  105014}, \href{http://arxiv.org/abs/1503.07537}{{\tt
  arXiv:1503.07537 [hep-ph]}}.

\bibitem{Lehman:2015coa}
L.~Lehman and A.~Martin, ``{Low-derivative operators of the Standard Model
  effective field theory via Hilbert series methods},''
  \href{http://dx.doi.org/10.1007/JHEP02(2016)081}{{\em JHEP} {\bf 02} (2016)
  081}, \href{http://arxiv.org/abs/1510.00372}{{\tt arXiv:1510.00372
  [hep-ph]}}.

\bibitem{Henning:2015alf}
B.~Henning, X.~Lu, T.~Melia, and H.~Murayama, ``{2, 84, 30, 993, 560, 15456,
  11962, 261485, ...: Higher dimension operators in the SM EFT},''
  \href{http://dx.doi.org/10.1007/JHEP08(2017)016}{{\em JHEP} {\bf 08} (2017)
  016}, \href{http://arxiv.org/abs/1512.03433}{{\tt arXiv:1512.03433
  [hep-ph]}}. [Erratum: JHEP 09, 019 (2019)].

\bibitem{Henning:2017fpj}
B.~Henning, X.~Lu, T.~Melia, and H.~Murayama, ``{Operator bases, $S$-matrices,
  and their partition functions},''
  \href{http://dx.doi.org/10.1007/JHEP10(2017)199}{{\em JHEP} {\bf 10} (2017)
  199}, \href{http://arxiv.org/abs/1706.08520}{{\tt arXiv:1706.08520
  [hep-th]}}.

\bibitem{Henning:2019mcv}
B.~Henning and T.~Melia, ``{Conformal-helicity duality \textbackslash{}\& the
  Hilbert space of free CFTs},'' \href{http://arxiv.org/abs/1902.06747}{{\tt
  arXiv:1902.06747 [hep-th]}}.

\bibitem{Henning:2019enq}
B.~Henning and T.~Melia, ``{Constructing effective field theories via their
  harmonics},'' \href{http://dx.doi.org/10.1103/PhysRevD.100.016015}{{\em Phys.
  Rev. D} {\bf 100} (2019) no.~1, 016015},
  \href{http://arxiv.org/abs/1902.06754}{{\tt arXiv:1902.06754 [hep-ph]}}.

\bibitem{Li:2020gnx}
H.-L. Li, Z.~Ren, J.~Shu, M.-L. Xiao, J.-H. Yu, and Y.-H. Zheng, ``{Complete
  set of dimension-eight operators in the standard model effective field
  theory},'' \href{http://dx.doi.org/10.1103/PhysRevD.104.015026}{{\em Phys.
  Rev. D} {\bf 104} (2021) no.~1, 015026},
  \href{http://arxiv.org/abs/2005.00008}{{\tt arXiv:2005.00008 [hep-ph]}}.

\bibitem{Li:2020tsi}
H.-L. Li, Z.~Ren, M.-L. Xiao, J.-H. Yu, and Y.-H. Zheng, ``{Low energy
  effective field theory operator basis at d \ensuremath{\leq} 9},''
  \href{http://dx.doi.org/10.1007/JHEP06(2021)138}{{\em JHEP} {\bf 06} (2021)
  138}, \href{http://arxiv.org/abs/2012.09188}{{\tt arXiv:2012.09188
  [hep-ph]}}.

\bibitem{Li:2020xlh}
H.-L. Li, Z.~Ren, M.-L. Xiao, J.-H. Yu, and Y.-H. Zheng, ``{Complete set of
  dimension-nine operators in the standard model effective field theory},''
  \href{http://dx.doi.org/10.1103/PhysRevD.104.015025}{{\em Phys. Rev. D} {\bf
  104} (2021) no.~1, 015025}, \href{http://arxiv.org/abs/2007.07899}{{\tt
  arXiv:2007.07899 [hep-ph]}}.

\bibitem{AccettulliHuber:2021uoa}
M.~Accettulli~Huber and S.~De~Angelis, ``{Standard Model EFTs via on-shell
  methods},'' \href{http://dx.doi.org/10.1007/JHEP11(2021)221}{{\em JHEP} {\bf
  11} (2021)  221}, \href{http://arxiv.org/abs/2108.03669}{{\tt
  arXiv:2108.03669 [hep-th]}}.

\bibitem{Li:2021tsq}
H.-L. Li, Z.~Ren, M.-L. Xiao, J.-H. Yu, and Y.-H. Zheng, ``{Operator bases in
  effective field theories with sterile neutrinos: d \ensuremath{\leq} 9},''
  \href{http://dx.doi.org/10.1007/JHEP11(2021)003}{{\em JHEP} {\bf 11} (2021)
  003}, \href{http://arxiv.org/abs/2105.09329}{{\tt arXiv:2105.09329
  [hep-ph]}}.

\bibitem{Li:2022tec}
H.-L. Li, Z.~Ren, M.-L. Xiao, J.-H. Yu, and Y.-H. Zheng, ``{Operators for
  generic effective field theory at any dimension: on-shell amplitude basis
  construction},'' \href{http://dx.doi.org/10.1007/JHEP04(2022)140}{{\em JHEP}
  {\bf 04} (2022)  140}, \href{http://arxiv.org/abs/2201.04639}{{\tt
  arXiv:2201.04639 [hep-ph]}}.

\bibitem{Li:2023cwy}
X.-X. Li, Z.~Ren, and J.-H. Yu, ``{A complete tree-level dictionary between
  simplified BSM models and SMEFT (d $\leq$ 7) operators},''
  \href{http://arxiv.org/abs/2307.10380}{{\tt arXiv:2307.10380 [hep-ph]}}.

\bibitem{Song:2023jqm}
H.~Song, H.~Sun, and J.-H. Yu, ``{Complete EFT Operator Bases for Dark Matter
  and Weakly-Interacting Light Particle},''
  \href{http://arxiv.org/abs/2306.05999}{{\tt arXiv:2306.05999 [hep-ph]}}.

\bibitem{Li:2023wdz}
H.-L. Li, Z.~Ren, M.-L. Xiao, J.-H. Yu, and Y.-H. Zheng, ``{On-shell Operator
  Construction in the Effective Field Theory of Gravity},''
  \href{http://arxiv.org/abs/2305.10481}{{\tt arXiv:2305.10481 [gr-qc]}}.

\bibitem{Harlander:2023psl}
R.~V. Harlander, T.~Kempkens, and M.~C. Schaaf, ``{Standard model effective
  field theory up to mass dimension 12},''
  \href{http://dx.doi.org/10.1103/PhysRevD.108.055020}{{\em Phys. Rev. D} {\bf
  108} (2023) no.~5, 055020}, \href{http://arxiv.org/abs/2305.06832}{{\tt
  arXiv:2305.06832 [hep-ph]}}.

\bibitem{Harlander:2023ozs}
R.~V. Harlander and M.~C. Schaaf, ``{AutoEFT: Automated operator construction
  for effective field theories},''
  \href{http://dx.doi.org/10.1016/j.cpc.2024.109198}{{\em Comput. Phys.
  Commun.} {\bf 300} (2024)  109198},
  \href{http://arxiv.org/abs/2309.15783}{{\tt arXiv:2309.15783 [hep-ph]}}.

\bibitem{Frame1954TheHG}
J.~S. Frame, G.~de~B.~Robinson, and R.~M. Thrall, ``The hook graphs of the
  symmetric group,'' {\em Canadian Journal of Mathematics} {\bf 6} (1954)  316
  -- 324. \url{https://api.semanticscholar.org/CorpusID:124024352}.

\bibitem{doi:10.1142/0097}
W.-K. Tung, \href{http://dx.doi.org/10.1142/0097}{{\em Group Theory in
  Physics}}.
\newblock WORLD SCIENTIFIC, 1985.
\newblock
  \href{http://arxiv.org/abs/https://www.worldscientific.com/doi/pdf/10.1142/0097}{{\tt
  https://www.worldscientific.com/doi/pdf/10.1142/0097}}.
\newblock \url{https://www.worldscientific.com/doi/abs/10.1142/0097}.

\bibitem{Fulton1996}
W.~Fulton, \href{http://dx.doi.org/10.1017/CBO9780511626241}{{\em Young
  Tableaux: With Applications to Representation Theory and Geometry}}.
\newblock Cambridge University Press, Cambridge, 1996.
\newblock
  \url{https://www.cambridge.org/core/product/A7570B10D82AE7233E25E5D6F70A07B6}.

\bibitem{Drummond:2008vq}
J.~M. Drummond, J.~Henn, G.~P. Korchemsky, and E.~Sokatchev, ``{Dual
  superconformal symmetry of scattering amplitudes in N=4 super-Yang-Mills
  theory},'' \href{http://dx.doi.org/10.1016/j.nuclphysb.2009.11.022}{{\em
  Nucl. Phys. B} {\bf 828} (2010)  317--374},
  \href{http://arxiv.org/abs/0807.1095}{{\tt arXiv:0807.1095 [hep-th]}}.

\end{thebibliography}\endgroup

\end{document}